\documentclass[english,aps,prl,amsmath,amssymb,twocolumn,superscriptaddress]{revtex4}
\usepackage{epsfig}
\usepackage{bm}
\usepackage{bbold}
\usepackage{latexsym}
\usepackage{color}

\begin{document}
	\title{On the experimental verification of the uncertainty principle of\\ position and momentum}
	
	\author{Thomas Sch\"urmann}
	\affiliation{J\"ulich Supercomputing Centre, J\"ulich Research Centre, D-52425 J\"ulich, Germany}
	\email{<t.schurmann@icloud.com>}	
	\author{Ingo Hoffmann}
	\author{Winfrid G\"orlich}
	\affiliation{Heinrich Heine University D\"usseldorf, 40225 D\"usseldorf, Germany}

	\date{ \today}
	
	\begin{abstract}
		Historically, Kennard was the first to choose the standard deviation as a quantitative measure of uncertainty, and neither he nor Heisenberg explicitly explained why this choice should be appropriate from the experimental physical point of view. If a particle is prepared by a single slit of spatial width $\Delta x$, it has been shown that a finite standard deviation $\sigma_p<\infty$ can only be ensured if the wave-function is zero at the edge of $\Delta x$, otherwise it does not exist\,\cite{TS09}. Under these circumstances the corresponding sharp inequality is $\sigma_p\Delta x\geq \pi\hbar$. This bound will be reconsidered from the mathematical point of view in terms of a variational problem in Hilbert space and will furthermore be tested in a 4f-single slit diffraction experiment of a laser beam. Our results will be compared with a laser-experiment recently given by Fernández-Guasti (2022) \cite{G22}.  
	\end{abstract}

\maketitle
\section{1.\,Introduction}
The diffraction of particles by a single slit has often been discussed as an illustration of Heisenberg's uncertainty relations and their role in the process of measurement. In the case of particles passing through a slit of width $\Delta x$ in a diaphragm of some experimental arrangement, the diffraction by the slit of the wave implies a spread in the momentum of the particles, which is greater the narrower the slit. This phenomenon is an example of the famous Heisenberg principle \cite{H27}\cite{H30}.
The most familiar formalization of the uncertainty principle is in terms of standard deviations \cite{K27}
\begin{eqnarray}\label{K}
	\sigma_x \sigma_p\geq \hbar/2
\end{eqnarray}
with the reduced Planck constant $\hbar=h/2\pi$. 

However, for particles passing through a slit of width $\Delta x$, the diffraction of the incoming wave function $\psi$ has to be considered as a preparation corresponding to the ordinary {\it von Neumann-L\"uders projection}. This approach is often applied in the actual experimental design of the uncertainty relation \cite{S69}\cite{Le69}\cite{Z03}. The advantage is that the localization of the particles is simply given by the width $\Delta x$ of the slit. Unfortunately this approach cannot be considered as a rigorous experimental test of Kennard's expression (\ref{K}) because $\Delta x$ and $\sigma_x$ are quite different measures of localization. An obvious interrelation between $\Delta x$ and $\sigma_x$ is given by the Popoviciu inequality for variances \cite{Po} 
\begin{eqnarray}\label{PO}
	\sigma_x \leq\frac{\Delta x}{2}    
\end{eqnarray}
while equality holds precisely when half of the probability is concentrated at each of the two boundary points of $\Delta x$. From (\ref{K}) and (\ref{PO}), the following inequality is immediately implied
\begin{eqnarray}\label{K1}
	\sigma_p \Delta x >  \hbar      
\end{eqnarray}
However, the bound $\hbar$ on the right-hand side can never be reached. This is because a finite value of $\sigma_p<\infty$
can only be ensured if the wave-function is zero at the edge of $\Delta x$, otherwise it does not exist. According to this condition the following sharp inequality has been introduced \cite{TS09}
\begin{eqnarray}\label{S}
	\sigma_p\Delta x\geq \pi\hbar.	
\end{eqnarray}
Since the standard deviation $\sigma_p$ is (by definition) corresponding only to the "one-sided" deviation of the probability density it is sometimes obvious to measure the momentum uncertainty $\Delta p$ according to the definition
\begin{eqnarray} \label{2s}
	\Delta p= 2\sigma_p.
\end{eqnarray}
With this definition, the inequalities (\ref{K1}) and (\ref{S}) can equivalently be rewritten as
\begin{eqnarray} \label{H0}
	\Delta x\Delta p > 2 \hbar        
\end{eqnarray}
and 
\begin{eqnarray} \label{H1}
	\Delta x\Delta p \geq 2 \pi \hbar. 
\end{eqnarray}
Actually, the sharp inequality (\ref{H1}) can be considered as a mathematically rigorous version of the heuristic expression $\Delta x\Delta p\gtrsim h$ originally introduced by the measurement process of Heisenberg in \cite{H27}\cite{H30}, before Kennard came with the alternative measurement approach corresponding to inequality (\ref{K}).  \\

Recently it has been mentioned that there is a quantum optical realization of the position and momentum uncertainty relation in the quantum limit \cite{G22}. Interference of two non-collinear photon modes with different frequencies are space and time resolved and the detection is performed in the same space-time region. It is argued that evaluation of the photon momentum from the position versus time interferograms makes this procedure similar to the mechanical momentum construct. The measured uncertainty $\Delta p_x$ in \cite{G22} is applied to approximate twice as much of the standard deviation $\sigma_p$ in order to verify the inequalities (\ref{H0}) and (\ref{H1}). The experimental factors corresponding to the right-hand side of (\ref{H0}) and (\ref{H1}) are given by $1.128\hbar$, $2.464\hbar $ and $2.723 \hbar$. That is, inequality (\ref{H0}) is experimentally falsified for the first measurement and expression (\ref{H1}) for all three cases. This conclusion seems at least questionable, such that a closer look at the measurement in \cite{G22} seems appropriate at this point. To do so, let us consider the experimental determination of $\Delta p_x$ given in \cite{G22} by
\begin{eqnarray} \label{4}
	\Delta p_x = \hbar\,\frac{\delta \omega}{2}\cdot\frac{\dot{x}_\text{max}-\dot{x}_\text{min}}{\dot{x}_\text{max}\,\dot{x}_\text{min}}.
\end{eqnarray}
The notation $\dot{x}=dx/dt$ is the velocity by which the fringes of the interferograms are displaced in time and $\delta\omega$ the frequency shift of the incoming modes expressed in Figs.\,3, 4 and 5 of \cite{G22}. In the experimental approach $\Delta p_x$ is considered to be proportional to the difference of nearby maximum and minimum slopes. Actually, the slopes extrema were drawn by hand prior to any calculation to avoid statistical bias. An arbitrary dark fringe was chosen and parallel slopes to the median were drawn (the distance between them was roughly half the distance between fringe maxima). The slopes extrema have been obtained from the diagonals of the corresponding parallelogram.  

As far as can be seen it is neither obvious nor proven in \cite{G22} that expression (\ref{4}) is an appropriate approximation of (\ref{2s}). The importance of this aspect is because the momentum uncertainties relevant in this context are exclusively based on the notion of statistical standard deviations (or twice of them). Actually, there is no remark in \cite{G22} why the difference of nearby maximum and minimum slopes justifies a proper approximation of $2 \sigma_p$. Only proportionality of (\ref{4}) and $\sigma_p$ has been mentioned, but this cannot be sufficient for the conclusion given in \cite{G22}.

In order to classify the measure of momentum uncertainty (\ref{4}) and the corresponding measurement results, let us first refer to a corollary of the measurement in Fernández-Guasti's work \cite{G22} (p.\,5):\\

\begingroup
\leftskip=0.5cm
\rightskip=0.5cm 
\noindent
"[...]\,The outcome of our experiments rule out this possibility $[\,\Delta x \Delta p_x \geq \pi\hbar\,]$, provided that our uncertainty estimates are equal or larger than the effective (overall) width of the distribution functions \cite{U85}, since all reported values in units of $\hbar$ are smaller than $\pi$.”\\
\par
\endgroup
\leftskip=0.0cm 
\noindent
The position uncertainty $\Delta x$ satisfies the "overall" width criterion mentioned in this corollary (for $N\to 1$ in \cite{U85}, Eq.(13)). \\
\\
However, according to the above corollary one also has to check if the momentum uncertainty estimate $\Delta p_x$ of the measurement in Ref.\,\cite{G22} is such that the probability 
\begin{eqnarray}\label{Prob0}
	P(\Delta p_x)=\int\limits_{-\Delta p_x/2}^{\Delta p_x/2}|\varphi(p)|^2\,dp
\end{eqnarray}
of the corresponding momentum density $|\varphi(p)|^2$ is sufficient close to 1 (for instance $\geq 70\%$). Otherwise, according to the criteria in Ref.\,\cite{U85}, the momentum uncertainty can not be considered to be well-defined. 

In order to check this criterion (which has not been discussed for the momentum uncertainty in Ref.\,\cite{G22}), we refer to a general {\it least upper bound} $\lambda_0$ of the probability expression (\ref{Prob0}), which is given by
\begin{eqnarray} \label{}
	\text{P}(\Delta p_x) \leq \lambda_0.
\end{eqnarray}
This bound follows from the work of Landau and Pollak (\cite{LP61},Theorem 2), Lenard (\cite{L72}, Proposition 11) or more recently in the context of quantum physics from Lathi \cite{L86} and Busch et\,al.\,\cite{BHL07}. 
The bound is a continuous and monotonically increasing function, $\lambda_0(\xi)$, of a single parameter given by
\begin{eqnarray} \label{xi}
	\xi = \frac{\Delta x \Delta p_x}{h},
\end{eqnarray}
with $\lambda_0(0)=0$ and $\lambda_0(\xi)\to 1$, for $\xi\to\infty$ (for details, see Fig.\,2 in \cite{LP61}, or Theorem 1 in \cite{S08}). \\
\\
Now, all measurement results given in Ref.\,\cite{G22}, (p.\,5), can be summarized as 
\begin{eqnarray} \label{ai}
	\Delta x\Delta p_x= a_i\hbar
\end{eqnarray}
for $i=1,2,3$ and 
\begin{eqnarray} \label{}
	a_1&=&1.128,\\
	a_2&=&2.464,\\
	a_3&=&2.723.
\end{eqnarray}
After equating expression (\ref{xi}) and (\ref{ai}), the $\xi$-values in (\ref{xi}) are given by $\xi_i=a_i/2\pi$, that is
\begin{eqnarray} \label{}
	\xi_1&=&0.179,\\
	\xi_2&=&0.392,\\
	\xi_3&=&0.433.
\end{eqnarray}
The numerical values of the corresponding upper bounds are given by
\begin{eqnarray} \label{}
	\lambda_0(\xi_1)&=&0.178,\\
	\lambda_0(\xi_2)&=&0.376,\\
	\lambda_0(\xi_3)&=&0.412.
\end{eqnarray}
In all cases, the probability weight of the momentum density (\ref{Prob0}) is far below $50\%$ and therefore $\Delta p_x$ in Ref.\,\cite{G22} cannot be considered as a reasonable measure of uncertainty with respect to criteria in \cite{U85}. It follows that the conditions for the applicability of the measurement corollary are not satisfied. \\

In contrast, a reasonable upper bound $\lambda_0(\xi)=78\%$ is obtained for $\xi = 1$ or equivalently $a = 2\pi$, such that instead of the expression  $"\Delta x\Delta p_x\sim 2 \hbar"$ mentioned in the abstract of Ref.\,\cite{G22}, equation (\ref{ai}) implies
\begin{eqnarray} \label{}
	\Delta x\Delta p_x= 2 \pi\hbar.
\end{eqnarray}
This expression is corresponding to our inequality (\ref{H1}) mentioned above. Finally, a correction on the right-hand side in the definition (\ref{4}) by a prefactor of $\pi$ might be appropriate. 
\\

In the following section a mathematical proof of our main inequality (\ref{S}) is given in terms of a variational problem in Hilbert space and the state of minimum uncertainty is derived. In Sec.\,3 an experimental setup for the verification of inequality (4) is presented. An approximated state of minimal uncertainty is prepared in a 4f-laser experiment with the help of a Lanczos-window to cut off higher spatial frequencies. From the data of the interference pattern, the momentum distribution and the corresponding momentum standard deviation $\sigma_p$ is determined. A discussion and conclusion is given in Sec.\,4.

\section{2.\, Variation problem in Hilbert space}
Let us consider particles in one spatial dimension described by a wave function $\psi$ which is an element of the Hilbert space ${\cal H}=L^2(\mathbb{R})$, the space of square integrable functions on $\mathbb{R}$. The scalar product in Hilbert space will be denoted by angular brackets, that is to write $\langle \phi|\psi\rangle$ for the scalar product of two state vectors $\phi,\psi\in{\cal H}$. Accordingly, the norm of $\psi$ is given by $||\psi||\equiv \sqrt{\langle \psi|\psi\rangle}$. The point-spectrum of the momentum eigenvalue equation 
\begin{eqnarray} \label{}
\hat p|\varphi_n\rangle =p_n|\varphi_n\rangle 
\end{eqnarray}
on the interval $[-\frac{\Delta x}{2},\frac{\Delta x}{2}]$ of length $\Delta x$ is given by 
\begin{eqnarray} \label{}
	\varphi_n(x) &=& \frac{1}{\sqrt{\Delta x}}\,e^{\frac{i}{\hbar} p_n x}\\
	p_n&=& \hbar \,k_n\\
	k_n&=&\frac{2 \pi n}{\Delta x}  
\end{eqnarray}
for $n=0,\pm 1,\pm 2,...$ This complete set of solutions satisfies the orthonormality relations  
\begin{eqnarray} \label{}
	\langle\varphi_n|\varphi_m\rangle = \delta _{nm}.
\end{eqnarray}
Thus, the decomposition  
\begin{eqnarray} \label{F}
	|\psi\rangle = \sum_{n=-\infty}^{\infty} c_n \,|\varphi_n\rangle 
\end{eqnarray}
with Fourier coefficients
\begin{eqnarray} \label{}
	c_n=\langle\varphi_n|\psi\rangle \in \mathbb{C}  
\end{eqnarray}
satisfies Parseval's theorem $||\psi||^2=\sum_n |c_n|^2$. Accordingly, the first and second moments of the momentum can be expressed by
\begin{eqnarray} 
	\langle \hat p\rangle &=& \frac{2\pi \hbar}{\Delta x} \,\sum_n n |c_n|^2 \label{p}\\
	\langle \hat p^2\rangle &=& \left(\frac{2\pi \hbar}{\Delta x}\right)^2 \sum_n n^2 |c_n|^2
\end{eqnarray}
such that $\sigma_p^2=\langle \hat p^2\rangle - \langle \hat p\rangle^2$ can be written as 
\begin{eqnarray} \label{}
	\sigma_p^2 = \left(\frac{2\pi \hbar}{\Delta x}\right)^2 \left[\sum_n n^2 |c_n|^2-\left(\sum_n n |c_n|^2\right)^2 \right].	 
\end{eqnarray}
The associated variational problem with $||\psi||=1$ and the necessary condition $\psi(\pm\frac{\Delta x}{2})=0$ is given by 
\begin{eqnarray} \label{}
	&&\sigma_p^2\to \min\\
\nonumber	\\
	&&\sum_n |c_n|^2=1  \label{c1} \\
	&&\sum_n (-1)^n c_n^*=0.  \label{c2}
\end{eqnarray}
This can be solved by the definition of a Lagrange function with two integrating factors $\alpha$ and $\beta$ of the following form
\begin{eqnarray} \label{c}	
	L = \sigma_p^2- \beta  \sum_n |c_n|^2 - \alpha \sum_n (-1)^n c_n^*.
\end{eqnarray} 
The variation of $c_n$ and $c_n^*$ is independent of each other and the condition $dL=0$ is performed with respect to $c_n^*$. After a few algebraic steps the resulting condition is given by
\begin{eqnarray} \label{}	
	\left[\left(\frac{2\pi \hbar}{\Delta x}\right)^2 n^2- \frac{4\pi \hbar}{\Delta x}\langle \hat p\rangle n -\beta\right] c_n =  (-1)^n \alpha
\end{eqnarray} 
for $n=0,\pm 1,\pm 2,...$ This infinite set of conditions for the determination of $c_n$ is non-linear because of the contribution of $\langle \hat p\rangle$ in the second term of the left-hand side. Up to a constant phase, the solution of this equation is given by
\begin{eqnarray} \label{c3}	
	c_n = \frac{\sqrt{8}\,\,}{\pi} \frac{(-1)^n}{1-4\,n^2}.
\end{eqnarray} 
By substitution it can be verified that the restrictions (\ref{c1}) and (\ref{c2}) are satisfied. Moreover, the symmetry $c_n=c_{-n}$ of the solution implies that $\langle \hat p\rangle=0$. This confirms the assumption made by the authors in \cite{TS09}. The momentum distribution $|c_n|^2$ of (\ref{c3}) corresponds to the equal sign in (\ref{S}) and (\ref{H1}). The associated position representation of the wave-function is obtained by Fourier transformation (\ref{F}) and given by
\begin{eqnarray} \label{f}	
	\psi(x) = \sqrt{\frac{2}{\Delta x}} \cos\left(\frac{\pi x}{\Delta x} \right)
\end{eqnarray} 
for all $x\in [-\frac{\Delta x}{2},\frac{\Delta x}{2}]$.
From the coefficients in (\ref{c3}) the corresponding continuous function is given by substitution of $k_n=2 \pi n/\Delta x$. In this limit the normalized $k$-state function is  
\begin{eqnarray} \label{k}	
	\tilde\psi(k) = 2\sqrt{\pi \Delta x}\,\,  \frac{\cos\left(\frac{\Delta x}{2} k \right)}{\pi^2-\Delta x^2 k^2}
\end{eqnarray} 
for every $k\in \mathbb R$. The standard deviation $\sigma_k$ of the wave-numbers $k$ can be obtained by ordinary integration and is given by $\sigma_k=\pi/\Delta x$. After multiplication with $\hbar$ this reproduces the equal sign of the main relation given in (\ref{S}). Both (\ref{f}) and (\ref{k}) are the state of minimum uncertainty corresponding to $\sigma_p\Delta x = \pi\hbar$.

\section{3.\, Experimental verification }

\begin{figure}[ht]
\scalebox{1.0}{	\includegraphics[width=9.3cm,height=7.0cm]{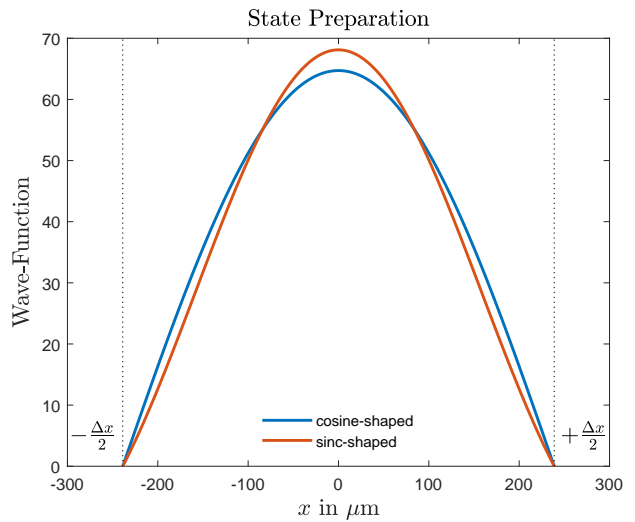}}
	\caption{State (\ref{f}) of minimum uncertainty (blue) and its Lanczos-window approximation (orange) given in (\ref{f1}). } \label{fig1}
\end{figure}

For an experimental verification of the lower bound in inequality (\ref{S}), a preparation of the state of minimal uncertainty (\ref{f}) would be necessary. Since a cosine-shaped wave function on a finite domain is a superposition of infinitely many partial waves, it is not obvious how this can experimentally be realized.	
However, an approximated state of minimal uncertainty can be prepared by apodization techniques in terms of a Lanczos-window to cut off higher spatial frequencies. The corresponding normalized sinc-function is given by   
\begin{eqnarray} \label{f1}	
	\phi(x) =\sqrt{\frac{\pi}{\text{Si}(2\pi) \Delta x}}\,\, \text{sinc}\left(\frac{2\pi }{\Delta x}\, x \right)
\end{eqnarray} 
with sinc$(x)=\frac{\sin x}{x}$ and Si$(x)=\int_0^\infty \sin(t)/t\, dt$. This function can be straightforwardly  prepared by plane-wave diffraction, see Fig.\,\ref{fig1}.
\noindent 
The exact value of the corresponding momentum standard deviation $\sigma_p$ is given by  
\begin{eqnarray} \label{f2}	
	\sigma_p \Delta x  = \gamma\, \pi \hbar
\end{eqnarray} 
where the constant $\gamma$ is  
\begin{eqnarray}
	\gamma &=& \frac{2}{\sqrt{3}\,}\, \sqrt{1-\frac{1}{\pi \, \text{Si}(2 \pi )}}\\
	&=& 1.0168880... \label{f3}	
\end{eqnarray} 
From the experimental point of view this numeric value is sufficiently close to 1, such that (\ref{f1}) can be considered as an appropriate approximation of the true state of minimum uncertainty in (\ref{f}). The corresponding normalized $k$-state function can be computed by Fourier integration and its density is  given by 
\begin{eqnarray} \label{k1}	
	|\tilde\phi(k)|^2 = \frac{\Delta x}{8\pi^2\text{Si}(2 \pi)} \left[\text{Si}\left(\frac{\Delta x}{2}k+\pi\right)-\text{Si}\left(\frac{\Delta x}{2}k-\pi\right) \right]^2\nonumber\\ 
\end{eqnarray} 
The corresponding diffraction pattern $I(y)$ at the screen of focal distance $f$ can be obtained in the ordinary Fresnel approach with $y/f=k/k_0$, where $y$ is the position on the screen. With (\ref{k1}), the intensity can be written as
\begin{eqnarray} \label{k2}	
	I(y) = \frac{k_0}{f}\, \left|\tilde\phi\left(\frac{k_0}{f} y\right)\right|^2
\end{eqnarray}    
where $k_0=2\pi/\lambda_0$. \\ 
\begin{figure}[th]
	\includegraphics[width=9.5cm,height=7cm]{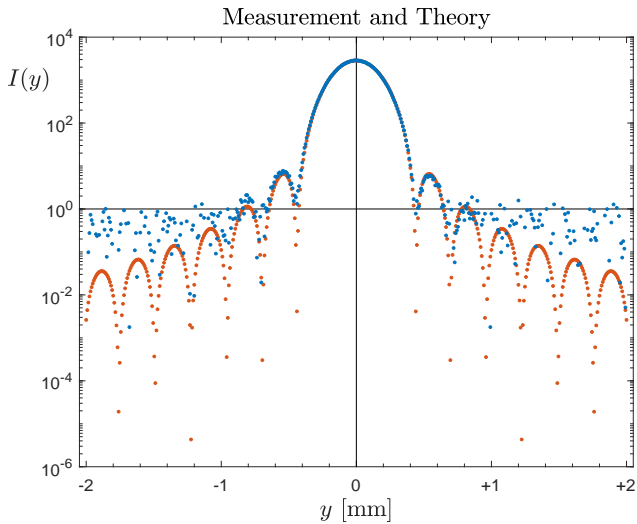}
	\caption{Measured (blue) versus theoretical (orange) density in semi-logarithmic representation. The signal-to-noise ratio outside the range of $y_c=\pm 1$\,mm is almost zero and therefore of minor relevance for the computation of expectation values under consideration.  } \label{fig2}
\end{figure}
In order to measure this intensity pattern, we use a HeNe monochromatic laser beam of wavelength $\lambda_0=\text{632.82\ n}$m (power: $\text{0.5 m}$W). The intensity of the beam is controlled by a polarization filter and subsequently collimated by the first beam expander seen in Fig.\,\ref{fig4}. After diffraction of the resulting plane-wave by the slit\,1 of fixed width $\Delta x_0$ the sinc-function amplitude is obtained immediately afterwards the beam has passed the beam expander 2
at focal distance $f$, where a variable slit\,2 is used to select the correct width $\Delta x$ of the Lanczos-window. Here, a careful adjustment of the slit width $\Delta x$ is made to ensure that the main peak is truncated precisely at the first two zeroes. This truncation is the crucial preparation of the beam (not the measurement). The intensity of the final diffraction pattern behind the beam expansion 3 is measured by a CCD-Linear-Sensor (Eureca TCD1304 16 Bit A/D). The measurement range and its resolution is given by $N=3648$ pixels with pixel-size $\delta y=8\,\mu $m. 
The measurement was taken over the entire range of the detector ($N\times\delta y=29.184$\,mm), Fig.\,\ref{fig2}
shows the relevant domain of the interference pattern. 
Due to the extremely fast decay of the momentum density (\ref{k1}), the signal-to-noise ratio outside the range of $y_c=\pm 1$\,mm is expected to be almost zero, such that the tails are only of very limited usefulness for the determination of the relevant quantities. However, the situation is not as simple as it seems, because the very fast decrease of the density corresponds on the other hand with zeros which show up symmetrically near the origin. 
To get an impression for the convergence of the relevant quantity, the numerical expectation value of $\gamma$ was determined step-by-step starting from the center of the CCD accumulatively according to 
\begin{eqnarray} \label{k3}	
	\hat\gamma_n = \frac{2\Delta x}{\lambda_0 f}\,\left[\sum_{i=\frac{N}{2}-n+1}^{\frac{N}{2}+n} \delta y\,\,y_i^2\,\hat I_i\right]^\frac{1}{2}
\end{eqnarray}   
for $n=1,2,...,N/2$, and $\hat I_i$ is the measured (normalized) intensity contribution given by the data at position $y_i$ of the screen. The convergence of the estimator $\hat\gamma_n$ is shown in Fig.\,\ref{fig3}. The slight overestimation of the theoretical prediction in the range of $0.7$\,mm to $2.5$\,mm is due to the zeros in the true density, which cannot be adequately represented by the finite resolvability of the CCD. Above $2.5$\,mm, the non-zero signals are so rare that this overestimation is no longer important. Therefore, we assume that beyond $2.5$\,mm only symmetrical noise plays a role which should not fundamentally question the results achieved up to that point. Nevertheless, no statistical significance test seems to be needed to confirm the hypothesis that $\gamma$ is greater than 1. An extension of the integration range would only lead to a summation of symmetric Gaussian noise, which, as expected, would not lead to any additional improvement. 
\begin{figure}[t]
	\includegraphics[width=9.5cm,height=7cm]{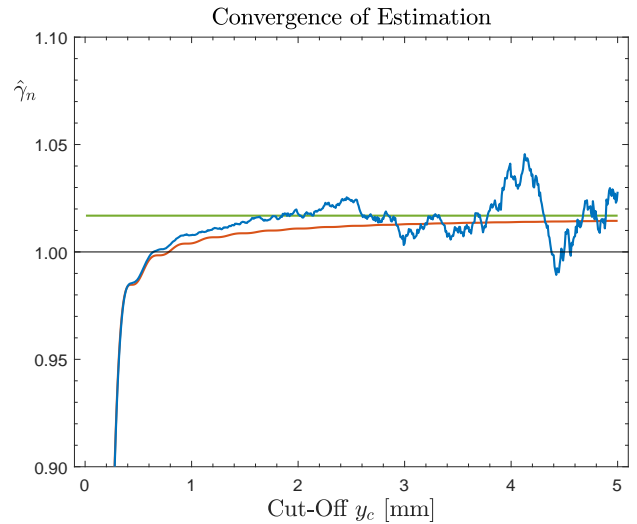}
	\caption{Stochastic convergence of the $\gamma_n$-estimator (\ref{k3}) (blue) and the exact value $\gamma=1.0168$ (green) according to (\ref{f3}). The theoretical value of the incremental expectation value according to (\ref{k1}) is monotonically approaching to $\gamma$ (orange).    } \label{fig3}
\end{figure}
\section{4.\, Summary and conclusion}

In this work, the importance of carefully choosing observables in experimentally testing Heisenberg's uncertainty principle has been emphasized. It has been discussed that a careless or often pragmatic approximation of standard deviations can lead to misinterpretations of experimental measurements and their implications. Since the popular uncertainty relations are mostly based on the notion of standard deviations, a closer look at their experimental determination is inevitable.\\ 

The exact determination of a standard deviation requires in most cases the integration over an infinite domain of a density function. In our 4f-setup, an experimental preparation  in terms of a Lanczos-window has been proposed such that the associated momentum density is sufficiently narrow to allow a reasonable experimental verification of our main inequality (\ref{S}).\\
  
Actually, this conclusion is not very surprising if one considers that the rigorous mathematical derivation of inequality (\ref{S}) in Sec.\,2. or in \cite{TS09} is based on the quantum mechanical foundations in Hilbert space and is by no means only a heuristic argument. Finally, it should be mentioned that the concept leading to the derivation of inequality (4) has been generalized to 3-dimensional manifolds with constant (nonzero) curvature in \cite{TS18}\cite{TS20}. 

\begin{figure*}[b]
	\includegraphics[width=18cm,height=6cm]{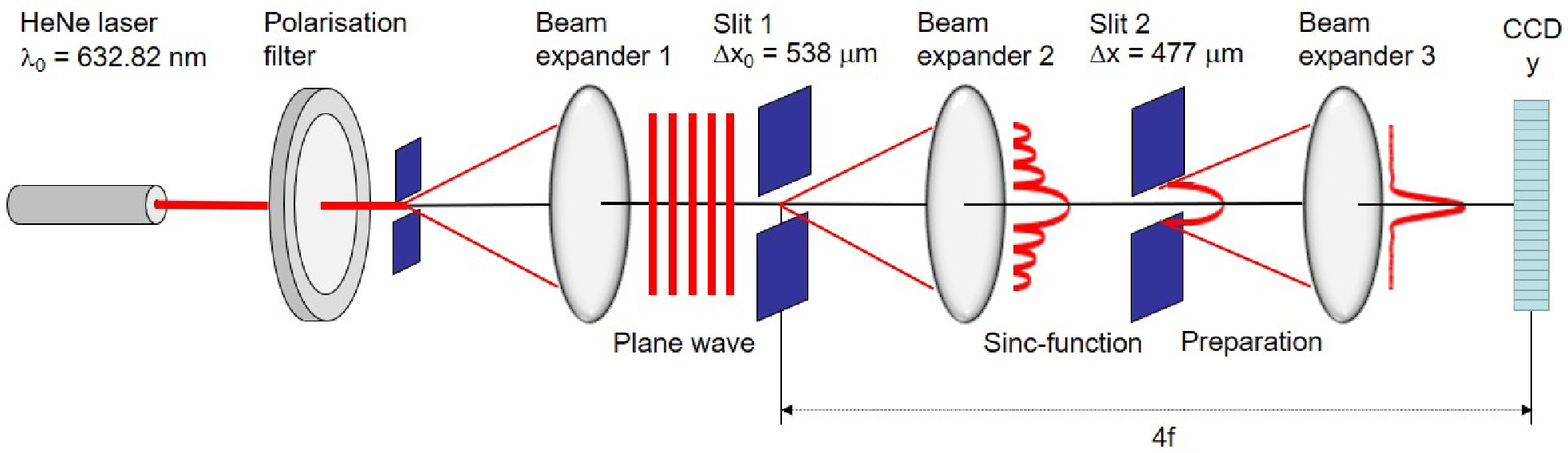}
	\caption{Schematic diagram of the experimental 4f-preparation. The expanded beam beyond the first lens (plane-wave) is modulated by two slits that single out the main-peak of a sinc-function by $\Delta x=477\,\mu$m (preparation). The intensity pattern corresponding to the minimum-uncertainty momentum density is detected by the CCD (measurement).} \label{fig4}
\end{figure*}

\newpage

\end{document}